\documentstyle[11pt,newpasp,twoside,epsf]{article}
\begin{document}

\title{Cosmic Microwave Background at low frequencies}

\author{Ravi Subrahmanyan}
\affil{Australia Telescope National Facility, CSIRO, Locked bag 194, 
Narrabri, NSW 2390, Australia}

\begin{abstract} 
The next generation low-frequency radio telescopes 
may \hfil\break probe cosmological 
models by means of observations of the cosmic microwave background (CMB).
I discuss the prospects for observations of CMB imprints --- recombination
lines from the epoch of recombination, $\mu$ distortions and angular 
temperature anisotropies --- at low frequencies.  
A future low-frequency radio telescope,
like the proposed SKA, may be capable of attempting some difficult
CMB measurements because of the large collecting area and large 
element numbers; however, this will require a telescope design that will
allow specialized calibration strategies and will give emphasis to
the control of spurious responses.
\end{abstract}

\section{Introduction}
The coming decade may well see a monumental enhancement 
in observing capabilities at low radio frequencies if, for example, 
the proposed
square-kilometre array (SKA) is constructed.  The quantum leap
in the total collecting area and in the numbers of antenna elements
constituting the array could move certain key observational
proposals from the realm of `dreams' to reality.  In this context --- and
while the telescope configuration and antenna element design are still
being debated --- it is perhaps timely that key and challenging
experiments that may be of relevance to the emerging opportunity 
are highlighted so that the instrument specifications may be tailored to 
eventually allow their endeavor.

The discovery of the existence of the cosmic microwave background (CMB),
followed by refinements in measurements of its characteristics
leading to the precise $COBE$ observations of its spectrum
and anisotropies, have been a major constraint and discriminant between
models of cosmology and structure formation.  Almost all measurements
that have been valuable in constraining  cosmology theories have been
made to-date at cm, mm and sub-mm wavelengths.  The long wavelength
measurements  of the CMB have been plagued by large errors owing to
the bright Galactic background temperatures and high levels of
extragalactic foreground discrete source confusion.

In this article, I shall attempt to highlight some aspects of CMB research
and discuss their relevance to low-frequency radio astronomy in the
context of the changed scenario expected in the coming decade if the
SKA is constructed.   Possible observing strategies and calibration
schemes are introduced; implications for the specifications for
the SKA are touched upon.

\section{The CMB temperature at low frequencies}

The $COBE-FIRAS$ experiment (Fixen et al. 1996) measured the temperature
of the cosmic microwave background (CMB) in the frequency range
70--640~GHz: no significant distortions from a Plankian form were detected 
and the best-fit thermodynamic temperature of the cosmic background
was determined to be $2.728 \pm 0.004$~K.  

Within the framework of the hot big bang cosmology, structure in the
universe is believed to have formed via gravitational instabilities
from primordial `seed' density perturbations.  The damping of sub-horizon
scale pressure waves --- as perturbations in the radiation field enter
the horizon over the redshift interval about $5 \times 10^{6}$ to
$5 \times 10^{5}$ --- are expected to {\it inevitably} leave their
imprint as a $\mu$-distortion in the CMB (Daly 1991; Hu, Scott, \& Silk
1994).  The wavelength $\lambda_{max}$ at which maximum distortion 
occurs is approximately 
$\lambda_{max} \simeq 2.2 (\Omega_{b} h^{2})^{-2/3}$~cm (Burigana, Danese,
\& De Zotti 1991); for $\Omega_{b} h^{2} \simeq 0.019$ (Burles, Nottel \&
Turner 1999), we may expect the temperature distortion to be a 
maximum at 30~cm wavelength.  The value of the chemical potential
$\mu_{\circ}$ --- as a consequence of the damping of primordial
density perturbations --- may be as small as $10^{-8}$ if the index
of the $COBE$-$DMR$ normalized matter power spectrum has an index $n=1$,
but could be as large as $10^{-4}$ if $n \simeq 1.6$ (Hu, Scott \& Silk
1994): correspondingly,
the maximum temperature distortion may be as large as 0.01~K.
Separately, any release of radiant energy in the redshift interval 
$5 \times 10^{6} > z > 5 \times 10^{5}$, perhaps owing to the decay
of particles with half lives in this range of cosmic times, 
would also result in $\mu$ distortions (Silk \& Stebbins 1983).

The $COBE$-$FIRAS$ measurements of the CMB spectrum 
limit $\mid \mu \mid$ to be less than $9 \times 10^{-5}$ (Fixen et al. 1996).
This implies that we may not expect a deviation exceeding about 0.008~K
at metre wavelengths as a consequence of any $\mu$ distortion.
Interestingly, these constraints placed on $\mu$ by the extremely precise
$COBE$-$FIRAS$ measurements are at least an order of magnitude more
valuable than the results from long-wavelength measurements of the
CMB spectrum, although they may have been made at frequencies where the
deviation may be a maximum.  

Recent measurements of the CMB temperature at frequencies close to 1~GHz
include the 600~MHz estimate of $T_{CMB} = 3.0 \pm 1.2$~K by 
Sironi et al. (1990), 
the 1400~MHz estimate of $T_{CMB} = 2.65 \pm 0.3$~K by 
Staggs et al. (1996) and 
the 1470~MHz estimate of $T_{CMB} = 2.26 \pm 0.19$~K by 
Bensadoun et al. (1993).  
A more recent estimate, albeit with larger errors, is that 
$T_{CMB} = 3.45 \pm 0.78$~K at 1280~MHz (Raghunathan \& Subrahmanyan 2000).

The large uncertainties in long-wavelength measurements of the absolute
brightness of the CMB are due, in part, to the relatively
high brightness of the Galactic background at these wavelengths,
the larger size of the receiving elements, the
greater difficulty in cooling them to cryogenic temperatures and consequently
the greater contribution from losses in the antenna and associated
feed.  The ground-based measurements also suffer from uncertainty
associated with estimations of the atmospheric contribution.
Improvements in estimates of $T_{CMB}$ at low frequencies may come
from improved methods of reducing losses
and/or developing methods of cancelling unwanted contributions, making
multifrequency measurements over an extremely wide frequency range, and
by placing the apparatus above the atmosphere.  

In this context it may be
mentioned that the experimental setup of  Raghunathan \& Subrahmanyan
used a novel technique of selecting cable lengths to cancel, via 
destructive interference,  an unwanted
contribution from the cold load connected to the third port of the circulator.
The $DIMES$ project (see the website at {\it ceylon.gsfc.nasa.gov/DIMES}),
with a wide spectral coverage from 2 to 100~GHz and based on a satellite 
platform, may be the kind of experiment that could 
improve sensitivity to $\mu$ distortions by more
than an order of magnitude.

\section{Fine structure in the CMB spectrum}

The cooling of the primeval plasma in the expanding universe is expected
to have led to recombination at a temperature near 3000~K (Peebles 1968;
Jones \& Wyse 1985).  It has been argued (Bartlett \& Stebbins 1991)
that measurements to-date of the CMB spectrum (that limit free-free
emission from an ionized intergalactic medium and $y$ distortions
arising from Compton scattering by hot electrons in such a medium)
do not require a neutral period and hence do not rule out the possibility
that the universe remained ionized throughout its history.  However,
the upper limits on the redshift of re-ionization (post recombination)
derived from the position of the peak in the spectrum of CMB
anisotropies (Griffiths, Barbosa, \& Liddle 1999) 
indicates that the universe was perhaps largely neutral
beyond a redshift of about 40: the data supports primordial recombination.
In this context, a direct observational probe of the recombination 
epoch would be valuable: the recombination lines that are {\it inevitably}
generated during recombination is a potential probe.  

The dominant additive recombination-line feature in the CMB spectrum is the 
L$_{\alpha}$ hydrogen line, this is expected to manifest in a spectral
`hump' at about 0.014~cm wavelength and a broad `continuum' extending 
to higher frequencies (Burdyuzha \& Chekmezov 1994).

Of interest to low-frequency radio astronomy are the hydrogen (and helium)
recombination lines
corresponding to transitions between highly excited states that may be
visible today at metre and centimetre wavelengths owing to the
extraordinary redshift ($z \sim 1100$) of the epoch of recombination.
As pointed out by Dubrovich \& Solyarov (1995) --- continuing on the earlier
work by Dubrovich (1975) and Bernshtein, Bernshtein, \& Dubrovich (1977) ---
the number of photons in transitions between adjacent levels 
as well as the ratio of the
distance between consecutive lines to the line widths both decrease with
increasing wavelength. As a consequence, the detection becomes extremely
challenging at low frequencies.  However, at least at cm wavelengths,
the extragalactic background light is dominated by the CMB whose
intensity varies inversely as the square of the wavelength; therefore,
the ratio of line strength to background continuum may be expected to
have a maximum at wavelengths 20--60~cm.  At these low frequencies,
the line is expected (Dell'Antonio \& Rybicki 1993; 
Dubrovich \& Solyarov 1995) to appear as 
spectral features with peak-to-peak brightness of about 0.1~$\mu$K.
The lines would be extremely broad because recombination is not 
`instantaneous': the redshift interval $\Delta z/z$ over which the
ionization fraction changes is about 20 per cent and is significantly
greater than that obtained assuming quasi-equilibrium (Saha) ionization
because recombination is `stalled' and `regulated' by the increased 
temperature in the Ly-$\alpha$ line as a consequence of recombination itself.
This results in the spectral lines manifesting as spectral ripples
with period 20 per cent of the observing frequency.

A detection of these spectral lines in the CMB would, apart
from clarifying the thermal history of the baryons, place constraints on the
baryon density and the mean matter density.
These spectral features are of extremely low intensity; however, if
radiant energy release in the early universe resulted in deviations in
the radiation background from the Planckian form, the spectral
features may have an increased prominence (Lyubarsky \& Sunyaev 1983).
In this case their observation may provide information on early energy
release.

\subsection{The observation of wideband CMB spectral features}

The detection of recombination lines or any other spectral features
that were added to the CMB in the early universe is an extremely
difficult and challenging experiment of the `high-risk high-gain' type.
It may be noted here that the spectral variations in temperature are
expected to be smaller than the angular anisotropy in the CMB.

Clearly, the measurements will have to be made with a `total-power'
type telescope; any interferometer would resolve the uniform sky signal.
Conventionally, `total-power' sky spectra are obtained from 
auto-correlations of the single-dish signals measured over a range of
delays.  However, the receiver-noise component may be eliminated from
the spectra by the use of `correlation receivers': this will require 
a built-in capability for splitting the signal from the feeds before
the front-end low-noise amplifiers.  Alternately, `total-power' sky
spectra may be measured using an interferometer array by 
scalar averaging the cross-power spectral amplitudes.  

\subsubsection{Sensitivity}

The detection requires extremely high brightness sensitivity:
the spectral features may be only 0.1~$\mu$K and a factor 
$10^{-8}$--$10^{-9}$ of the system temperature.  It is obvious that,
despite the large bandwidths involved, the signals cannot be detected
in any reasonable time using a single element telescope.
However, because the signals are
isotropic to a high degree, the antenna size is not directly
a determinant of the sensitivity.  Therefore, we may detect the signal
by averaging the sky spectra from a large number of receivers behind
relatively-small spatially-separated antenna elements. 
If we assume that integration
times of order 100~hr are realistic (very much larger times would make 
the debugging of the system difficult), and that signals are to be 
detected with spectral resolution of order 50~MHz at 1~GHz,  the number
of independent receivers is of order $10^{3}$ for a 3-$\sigma$ detection.
Perhaps the elements of an SKA may be designed to attempt this
experiment.

\subsubsection{Wide bandwidth}

A second cause for difficulty with this experiment is the wide bandwidth
required for the detection of these lines;
receivers with low noise temperatures over a wide bandwidth exceeding
20 percent of the observing frequency are required.  The receiver 
characteristics would undoubtedly change over this band and will have to
be calibrated; however, spurious broadband spectral features may appear
in the received spectrum as a result of any changes in the sky reception
pattern over the band.  Wide bandwidth observations, particularly at low
frequencies, are also susceptible to radio interference: the experiment
will require receivers with high dynamic range and interference
excising/cancelling techniques.

\subsubsection{Calibration}

Perhaps the greatest challenge would be the accurate calibration of the
instrument response.  Standard calibration techniques like beam switching
fail because the signals are isotropic.  Frequency switching may not be
useful because no spectral region is line-free. Ideally what one would 
like to do to calibrate the instrument is to make separate observations
with only the interesting signal `switched off'.

A possible calibration strategy is to make the CMB spectral measurements
using the individual elements of an array of telescopes, 
in `total power' mode, and `switch off' the uniform sky signal,
for the purpose of calibration, by operating the array as an interferometer.
The individual antenna spectral responses are determined from the
spectral visibilities obtained while
observing a strong, largely unresolved source, in interferometric mode.
Such an approach may also allow calibration measurements to be made 
simultaneously with the total-power CMB spectral measurements: time variations
in the band-pass characteristics would not then be a limiting factor.

In Fig.~1, I show an example where this calibration strategy has been
applied to spectra of Galactic H{\sc i}.  These spectra were obtained
with the Australia Telescope Compact Array (ATCA). Fig.~1(a) shows 
auto-correlation spectra obtained with the individual antennae,
Fig.~1(b) shows the scalar-averaged cross-power spectra obtained by taking
the spectral visibilities (these have been vector averaged on-line over 10~s;
within the narrow spectral bandwidth this averaging time is not
enough to detect the sources in the field with signal-to-noise exceeding
unity) on the interferometer baselines and averaging the amplitude spectra
off-line.  Fig.~1(c) shows the calibration spectrum obtained from the
spectral visibilities measured on an unresolved calibrator.  Fig.~1(d)
and (e) show, respectively, the calibrated auto-correlation 
and scalar-averaged cross-power spectra.  The signal-to-noise 
in the scalar-averaged
spectra are worse because the scalar averaging of amplitudes was done
only off line: the on-line 10~s averaging was performed vectorially 
in the individual 2-kHz channels and,
as a consequence, the amplitudes reduced by a factor 200.

\begin{figure}
\plotone{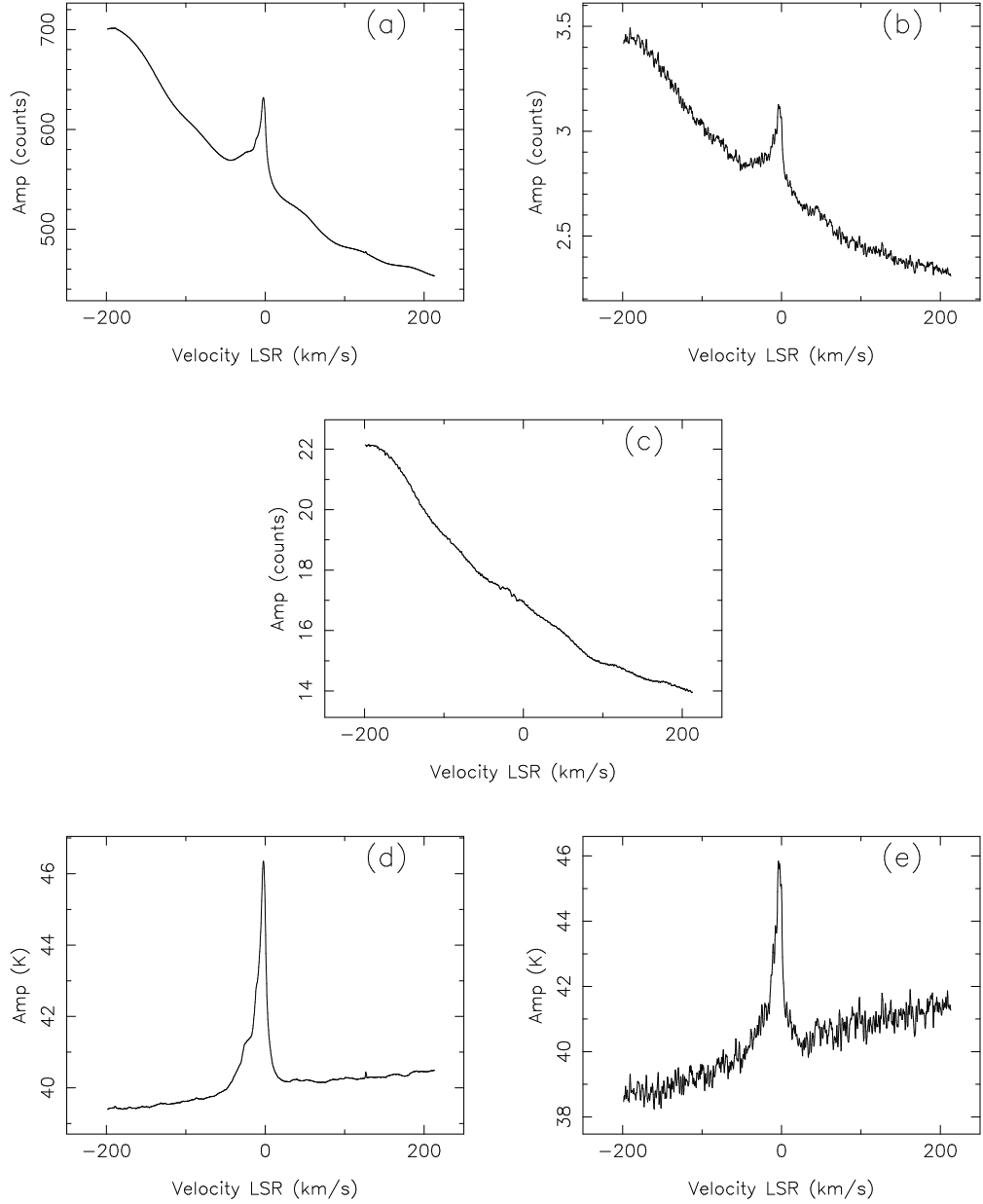}
\caption{ATCA spectra of Galactic H{\sc i}. (a) and (b) are, respectively, 
a single-dish auto-correlation spectrum and a scalar (amplitude) averaged 
cross-correlation spectrum.  (c) is the bandpass calibration determined from
interferometric observations of an unresolved calibrator.  (d) and (e) are
the corresponding calibrated spectra.}
\end{figure}

A disadvantage of such a scheme is that the calibration does not
include the receiver noise characteristics; however, if the total-power
spectra are obtained using correlation receivers, as discussed above,
the spectra will not include a contribution from receiver noise.
Another negative aspect of such an approach is that the calibration
will require a strong source: the antenna temperature due to the
calibrator source will have to be comparable to the system
temperature if the calibration time should be comparable to the
observing time.

Alternate methods of calibrating spectra, that may be worth exploring,
are (i) using the `Moon' to block the sky spectrum (Shaver et al. 1999;
Stankevich, Wielebinski, \& Wilson 1970) and (ii) equalizing the instrument
bandpass response by observing each sky spectral component through every 
spectrometer channel.  The former may impose severe requirements on the
linearity of the receivers and the latter method will require `fine-tuning'
capability in the first LO in the receiver chain.

\section{Angular anisotropies in the CMB temperature}

Most measurements of small-angle CMB anisotropy are being 
done at high radio frequencies (in the 15--90~GHz range)
where Galactic and extragalactic contaminants are a minimum.
However, the high-frequency detections of small-angle
anisotropy, which have to-date been Sunyaev-Zeldovich 
(S-Z) decrements towards
clusters of galaxies, have been difficult measurements made
with long integrations and using telescopes with small fields-of-view.
All-sky images of CMB anisotropies with 10's of arcmin resolution
are expected to become available in the coming decade from the
$MAP$ and $PLANCK$ satellite missions; however, they are expected to
detect S-Z anisotropies from
only the relatively high-mass and nearby S-Z clusters.
The next generation of high-brightness-sensitive imaging arrays,
like the CBI and AMIBA, which are specifically designed for surveys
for S-Z clusters, are also expected to cover only small sky areas
because of their small fields of view.

Owing to the upturn in the source counts at low flux density levels,
the $\mu$Jy source counts at GHz frequencies are approximately
(see, for example, Fig.~3 in Windhorst et al. 1993)

\begin{equation}
n(S)~=~10^{8}~S_{\mu{\rm Jy}}^{-2.2}
~f_{GHz}^{-0.8}~{\rm arcmin}^{-2}~{\rm Jy}^{-1}.
\end{equation}

\noindent In images made with a beam of FWHM $\theta$ arcmin,
these sources may be expected to result in a 1-$\sigma$ 
confusion rms given by

\begin{equation}
\Delta S~=~40~\theta^{1.7}~f_{GHz}^{-0.7}~\mu{\rm Jy}, ~{\rm or}~
\Delta T~=~14~\theta^{0.3}~f_{GHz}^{-2.7}~{\rm mK}.
\end{equation}

The arcmin-resolution anisotropy searches made at 
frequencies $< 10$~GHz --- including those with the
VLA and the ATCA --- have been limited by foreground discrete-source
confusion. In this context, it is useful to ask whether a future low-frequency
telescope, like the SKA, which will have the capability of imaging wide fields
of view, could make large-area surveys for S-Z clusters
and image the decrements with sufficient angular resolution to detect
any sub-structure.  The answer will depend on whether the enormous
improvement in flux sensitivity in an SKA-type telescope will enable 
surveys for low-surface-brightness S-Z clusters below the confusion limit
by detecting and subtracting a large part of the discrete-source confusion.

If we assume that all discrete sources above a lower flux density
limit of $S_m~\mu$Jy are subtracted from the sky images, the residual
confusion rms is

\begin{equation}
\Delta S~=~8~\theta~f_{GHz}^{-0.4}~S_m^{0.4}~\mu{\rm Jy}, ~{\rm or}~
\Delta T~=~3~\theta^{-1}~f_{GHz}^{-2.4}~S_m^{0.4}~{\rm mK}. 
\end{equation}

\noindent The proposed SKA is to have a sensitivity: 
$A_{eff}/T_{sys} = 2 \times 10^{4}$~m$^{2}$~K$^{-1}$.  We may expect
the continuum images to have a thermal noise of about 50~nJy with 10~hr
integration time.  Assuming that foreground sources above $S_m \sim 250$~nJy
are successfully subtracted, the residual confusion in arcmin resolution
images may be expected to be as large as 2~mK at 1~GHz but as low as 6~$\mu$K
at 10~GHz.

Clearly, any useful survey for S-Z clusters at low frequencies, even with
an SKA, will require very long integration times and observations at
frequencies $\ga10$~GHz.  However, because confusion will be an important 
limiting factor, it is important that particular attention be given to the
design of the antenna element and array configuration in order to
ensure that the sidelobe levels are low and roll off rapidly.

\section{Acknowledgments}
The Australia telescope is funded by the Commonwealth of Australia for 
operation as a National facility managed by CSIRO.  The HI spectra
displayed were obtained as part of collborative work with Mark Walker
of the UNSW, Australia.

\end{document}